\documentclass[a4paper,12pt]{article}

 \newcommand{\be}{\begin{equation}}
 \newcommand{\ee}{\end{equation}}
 \newcommand{\ii}{\'{\i}}
  \usepackage{amsmath,epsfig,amssymb,amsfonts,verbatim}
  \usepackage{fancyhdr}
  \usepackage[english]{babel}
  \usepackage{hyperref}

 \hypersetup{colorlinks,%
 filecolor=magenta,%
 citecolor=cyan,%
 linkcolor=blue,%
 urlcolor=magenta,%
 pdftex}

\begin{document}

 \thispagestyle{empty}

 \begin{center}
 \Huge{\bf  ON FACTS IN\\ SUPERSTRING THEORY}\footnote{ A detailed report on the sociological process here discussed is
available at
\href{http://spinningthesuperweb.blogspot.com}{Spinning the
Superweb}: Essays on the history of superstring theory
(www.spinningthesuperweb.blogspot.com). See
\href{http://spinningthesuperweb.blogspot.com/2008/05/1-on-facts-in-superstring-theory-iii.html}{section
III} and
\href{http://spinningthesuperweb.blogspot.com/2008/05/selected-readings-for-chapter-1.html}{IV}
of the first essay ``On Facts in Superstring Theory." These last
two sections are not included in this arXiv version!}

 \end{center}
  \begin{center}
  \Large{\bf Oswaldo Zapata Mar\ii n}
 \end{center}
 \vspace{.2cm}
  {\footnotesize {\bf Abstract:} Despite the lack of experimental confirmation and of unambiguous theoretical proof, superstring theory has long been considered by many the only consistent quantized theory of gravity and the unique viable framework for the unification of all fundamental forces of nature. In the first part of this essay I explore the type of reasoning used to support such statements. In order to illustrate the argument, in the second part I focus on one of the most acclaimed achievements of the theory: the AdS/CFT correspondence. Finally, I conclude by observing that what constitutes a result in superstring theory involves more than purely theoretical arguments. Specifically, the acceptance of facts in superstring theory is inextricably linked to the large group of people that make it possible, whether they are string practitioners or
  not.}

% \begin{quote}
% {\bf Abstract:}
% \end{quote}

 \section{The evolving scientific status of string theory results}
Trying to overcome the impasse with what a massless particle with
spin two meant for the dual model of strong nuclear interactions,
in 1974 Jo\"el Scherk and John Schwarz proposed a reinterpretation
of this particle as the quantum carrier of gravitational force:
``The possibility of describing particles other than hadrons
(leptons, photons, gauge bosons, gravitons, etc.) by a dual model
is explored. The Virasoro-Shapiro model is studied first,
interpreting the massless spin-two state of the model as a
graviton.''\footnote{ J. Scherk and J. H. Schwarz, ``Dual Models
for Nonhadrons,'' \emph {Nuclear Physics B} 81 (1974): 118-144.}
In their seminal paper Scherk and Schwarz showed that consistency
of the dual model entailed a higher dimensional version of the
Hilbert action. From this it then followed that the model included
gravitational forces as described by Einstein's equation. These
were the primary motives driving the authors to propose that
string theory quantized gravity: the spectrum showed a massless
spin-2 particle, and, moreover, a ten-dimensional Einstein
equation could be derived. In those days many theoretical
physicists found these two results too weak to allege that a
quantized theory of gravity had been achieved. This explains the
cautious reception the theory received in its early years.
Unexpectedly, however, this once feeble proposal has become widely
established within theoretical physics, even though the
mathematical support has remained almost unchanged for more than
thirty years. Let us comment further on this.

In the last chapter of a classic string theory graduate textbook
written in 1989, here is how the quantization of gravity is
presented:

\begin{quote}
String theory is claimed to be a unifying framework for the
description of all particles and their interactions, including
gravity. However, up to now our exposition of the subject was
rather formal and it is not at all transparent how it can be
relevant for low energy phenomenology. The only hint we got so far
was from looking at the spectrum. There especially the occurrence
of a spin two tensor particle indicated that gravity might be
contained in string theory.\footnote{ D. L\"ust and S. Theisen,
\emph{Lectures on String Theory} (Berlin: Springer-Verlag, 1989).}
\end{quote}

Note that this observation is relegated to the last part of the
book, after some three hundred pages of mathematical details. No
doubt this is a queer situation. Why leave the most important
argument in favour of the theory, at least in popular accounts and
undergraduate level materials, to the final pages of the textbook?
The answer to this question is provided by the cautious words of
the authors. Even more surprising is that the same argument has
been used for decades. The only difference is that nowadays the
quantization of gravity is not considered an ``elusive task,'' as
many string theoreticians used to say, but rather an accomplished
one. For example, in the midst of what string theorists consider
the second major revolution of the field, Edward Witten wrote in
\emph{Physics Today}: ``Moreover, these theories have (or this one
theory has) the remarkable property of \emph{predicting gravity}
-- that is, of requiring the existence of a massless spin-2
particle whose coupling at long distances are those of general
relativity.''\footnote{ E. Witten, ``Reflections on the Fate of
Spacetime,'' \emph{Physics Today}, April, 1996.} (Italics in the
original.) And in an up to date textbook, aimed at undergraduate
physics students, the author says: ``The striking quantum
emergence of gravitation in string theory has the full flavor of a
prediction.''\footnote{ B. Zwiebach, \emph{A First Course in
String Theory} (Cambridge: Cambridge University Press, 2004).}
Understandably, declarations of this kind have given rise to hot
discussions among supporters and detractors of the theory. The
question is: what happened in those intervening years? Why are
string theorists so optimistic now? Did they really find an
unquestionable proof, experimental or theoretical, that their
theory quantizes gravity?

In point of fact, to this day nobody has presented an entirely
convincing proof of this. For some theoretical physicists, the
presence of a massless spin-2 particle in its spectrum is not
enough to a quantized theory of gravity. I is also argued that the
low energy limit analysis of superstring theory does not imply
that that particle is the graviton. The former string theorist
Daniel Friedan, one of the early major contributors, is emphatic
about this:

\begin{quote}
In particular, there is no justification for the claim that string
theory explains or predicts gravity. String theory gives
perturbative scattering amplitudes of gravitons. Gravitons have
never been observed. Gravity in the real world is accurately
described by general relativity, which is a classical field
theory. There is no derivation of general relativity from string
theory. ... String theory does not produce any mechanical theory
of gravity, much less a quantum mechanical theory.\footnote{ D.
Friedan, ``A Tentative Theory of Large Distance Physics,'' [arXiv:
hep-ph/0204131].}
\end{quote}

Why then do practitioners believe that ``string theory is a
quantum theory, and, because it includes gravitation, it is a
quantum theory of gravity''? How have string theorists arrived at
the conclusion that ``the harmonious union of general relativity
and quantum mechanics is a major success''\footnote{ B. Greene,
\emph{The Elegant Universe: Superstrings, Hidden Dimensions, and
the Quest for the Ultimate Theory} (New York: W. W. Norton \& Co.,
1999).} of superstrings? Moreover, how have they managed to
convince other theoreticians of the validity of their
explanations? Let us look at another example.

In the abovementioned paper, Scherk and Schwarz also declared that
string theory could unify all the fundamental interactions: ``If
it is, a scheme of this sort might provide a unified theory of
weak, electromagnetic, and gravitational interactions.''\footnote{
J. Scherk and J. H. Schwarz, ''Dual Models for Nonhadrons,'' op.
cit.} (These known interactions were then complemented with the
strong nuclear force, the latter successfully described by quantum
chromodynamics.) This was in 1974. Years later, and after intense
work, the proposal had still not been proved. The four-dimensional
standard model (it exludes gravity), with all its details, could
not be deduced from string theory. In a lecture given at the
International Centre for Theoretical Physics (ICTP) in 1986, one
of the leading string phenomenologists of the time stated: ``Being
defined in d=10, some compactification of the six dimensions would
be required to make contact with phenomenology. This process is at
the moment not understood at all; one has to make crude
approximations and then check for consistency a
posteriori.''\footnote{ H. P. Nilles, ``Supergravity And The
Low-Energy Limit Of Superstring Theories,'' lectures given at the
1986 ICTP Spring School on Supersymmetry and Supergravity, in B.
de Wit et al. (eds.), \emph{Supersymmetry, Supergravity and
Superstrings 1986} (Singapore: World Scientific, 1986).}  The
reason why the process of superstring compactification ``was not
understood at all'' is due to the stringent constrains that
supersymmetry imposes on the four dimensional model. As expected
from the standard model of particle physics, any physical result
with supersymmetry must be renormalizable\footnote{ In broad
terms, this means that the theory cannot contain quantities which
take on an infinite value.}  and requires the existence of chiral
spinors.\footnote{ Chirality is an essential property of the
particles constituting matter: leptons and quarks.} However, there
are some difficulties with this: firstly, supergravity\footnote{
Supergravity is the theory known to be the low energy limit of
superstring theory.}  with one supersymmetry is not
renormalizable, and, secondly, models with higher numbers of
supersymmetries do not include chiral spinors. Satisfying these
two conditions is the difficult mission assigned to string
phenomenologists. There are several approaches to the problem: the
simplest model considers compactification on a multi-dimensional
torus, other string theorists prefer to use constructs known as
orbifolds or orientifolds;  more recently $G_2$ manifolds were
tested. Other Calabi-Yau manifolds are currently under
examination. Despite this confusing situation, there is one thing
string theorists know they must answer: ``Why do we live on this
particular string vacuum or SSC [superstring
compactification]?''\footnote{ F. Quevedo, ``Lectures on
Superstring Phenomenology,'' [arXiv: hep-th/9603074].}  This is
the most urgent question that needs to be addressed in order to
make contact with physical reality. As Michio Kaku writes in the
introduction to the 2000 edition of his textbook on elementary
string theory: ``The search for the true vacuum of string theory
is therefore the central theme of this book.''\footnote{ ``Thus,
the real problem facing us, in our opinion, is to theoretically
settle the following question as quickly as possible: what is the
vacuum (ground state) of superstring theory? Since the ground
state should correspond to our physical universe, if the true
vacuum could be discovered, we might be able to decisively settle
whether superstring theory is a theory of the universe or just the
latest in a series of failed efforts to discover the Holy Grail of
physics, the unified field theory.'' M. Kaku, \emph{Strings,
Conformal Fields, and M-Theory} (Berlin: Springer-Verlag, 2000),
second edition.}  So, if we could explain why the universe chose
this particular vacuum we would be able to understand how the
standard model arises from superstring theory and why the universe
expands as it does. It has been argued by critics that this
reformulation of the problem does not solve it. On the contrary,
it makes it harder and moves it, dangerously, towards the realm of
theology.\footnote{ By the year 2005 there was a vigorous debate
within theoretical physics concerning the role of anthropic
arguments in determining the real vacuum of the universe. Leonard
Susskind's \emph{The Cosmic Landscape} triggered this passionate
debate. The attacks on string theory during that period were
unprecedented. For example, Lawrence Krauss, a reputed
astrophysicist, wrote that ``it is perhaps not too surprising that
when one approaches the limits of our knowledge, theologians and
scientists alike tend to appeal to new hidden universes for,
respectively, either redemption or understanding.'' L. Krauss,
``Science and Religion Share Fascination in Things Unseen,''
\emph{The New York Times}, 8 November 2005. In summary, as Paul
Feyerabend would put it, are they scientists or priests?}

In this case, as in the previous case of the quantization of
gravity, superstring theorists have been unable to offer an
accurate and comprehensive explanation of four-dimensional
physics. To be sure, string phenomenology, from the old heterotic
string to recent brane models, does not provide the correct value
for the quantities associated to the elementary particles known so
far. In addition to this, critics emphasize, it does not answer
crucial questions that intrigue particle physicists: how is the
electroweak symmetry broken? what fixes the masses of the Higgs
boson, quarks, neutrinos and charged leptons? what are the sources
of the cold dark matter? what produced the big bang? why is there
matter-antimatter asymmetry? This state of affairs has lead
Sheldom Glashow, an eminent particle physicist, to declare that
string theory ``has failed in its primary goal, which is to
incorporate what we already know into a consistent theory that
explains gravity as well. The new theory must incorporate the old
theory and say something more. String theory has not succeeded in
this fashion.''\footnote{ Sheldom Glashow interview; on the
website of \emph{The Elegant Universe},
http://www.pbs.org/wgbh/nova/elegant/view-glashow.html }

From the previous examples we have learnt some important things
about the development of string theory. Firstly, as research
progresses in a given topic, an explicit reference to the
\emph{unsolved problem} tends to disappear from the literature.
For instance, we saw how the quantization of gravity is considered
by string theorists to be an accomplished task that does not
deserve further study, or even a mention. Secondly, while research
advances, the initial problem changes in such a way that it
becomes increasingly difficult to unravel the convoluted
relationship connecting the final problem to the original one.
This was illustrated by our second example concerning string
theory and the unification of the forces. Originally the idea was
to extract the standard model from superstring theory, an
investigation encouraged during the second half of the eighties by
the promising results obtained from the heterotic string. Then, by
the mid-nineties, the goal was to determine the unique vacuum of
the mother of all the theories, the M-Theory. And, more recently,
the focus was on the right ``environment'' of the anthropic
solution. Things have changed, but the fundamental query remains
unsolved: how do we get the standard model from string theory?
With these examples we have learnt something else: this occurs
while an ``outward'' discourse (from the ``inside'' to the
``outside'' of the professional community) proclaims that the
theory has solved such problems. Indeed, in this movement
disadvantages have been transmuted into virtues.

In spite of these fundamental flaws in the theory, enthusiasts
proclaim that ``in string theory all forces are truly unified in a
deep and significant way,'' or, a bit more prudently, ``string
theory leads in a remarkably simple way to a reasonable rough
draft of particle physics unified with gravity.''\footnote{ E.
Witten, ``Unravelling String Theory,'' \emph{Nature} 438,
December, 2005.} The final outcome of this discourse is the same:
the stabilization of string theory as a quantized theory of
gravity and unified model. Before concluding this introduction, I
would like to add two more quotations. In Zwiebach's undergraduate
textbook he asks:

\begin{quote}
Why is string theory truly a unified theory? The reason is simple
and goes to the heart of the theory. In string theory, each
particle is identified as a particular vibrational mode of an
elementary microscopic string.\footnote{ B. Zwiebach, \emph{A
First Course in String Theory}, op. cit.}
\end{quote}

Thus, string theory is a unified theory thanks to its extreme
reductionist approach. Brian Greene, in his best-selling book,
backs up this statement.  Years of hard work have shown that the
reductionist approach to string theory is correct.

\begin{quote}
These works showed conclusively that numerous features of the
standard model -- features that had been painstakingly discovered
over the course of decades of research -- emerged naturally and
simply from the grand structure of string theory.\footnote{ B.
Greene, \emph{The Elegant Universe}, op. cit.}
\end{quote}

I think what we have seen in these examples is a characteristic of
string theory research and its elaboration of physical reality. At
first, a hypothesis is made, explaining openly its significance as
well as its difficulties. At this stage no one is sure of the real
value of the conjecture, however, it is interesting enough to
drive a significant part of the physics community to devote itself
to its development. Step by step ``evidence'' accumulates and
after a while the string theory fact emerges. String theorists
have created in this way their own nature: a supersymmetric world,
a big bang with all the fundamental forces combined, a
multi-dimensional universe, and so forth. Although I have provided
support to this thesis in the case of the two sanctioned
``achievements'' of superstring theory, quantization of gravity
and unification of the fundamental forces, I will now illustrate
in full detail this complex process with another ground-breaking
proposal: the AdS/CFT correspondence.

 \section{A case study: the AdS/CFT correspondence}

At the end of 1997, Juan Maldacena, at that time a young
researcher at Harvard University, proposed what some physicists
consider to be one of the main breakthroughs in the history of
string theory and even of theoretical physics. Approaching the
physics of black holes with the powerful mathematical tools of
superstring theory, he conjectured the existence of a deep
relationship between pure non-gravitational theories and
superstring theories.\footnote{ J. M. Maldacena, ``The Large N
Limit of Superconformal Field Theory and Supergravity,'' [arXiv:
hep-th/9711200]. To put it simply, the correspondence says, as it
is currently understood, that string theory defined in a
negatively curved anti-de Sitter space (AdS) is equivalent to a
certain conformal field theory (CFT) living on its boundary. One
concrete example is AdS$_5$/CFT$_4$. It states that type IIB
superstring theory in AdS$_5$ is equivalently described by an
extended $N$=4 super-CFT in four dimensions. The other five
dimensions of the ten-dimensional space defined by superstring
theory are compactified on a five-dimensional sphere, S$^5$. The
five-sphere with isometry group SO(6) is chosen in order to match
with the SU(4) R-symmetry of the super Yang-Mills theory defined
on the boundary. More complicated models are also possible. Note
that there are two different N's involved in our discussion: $N$,
in italics, for the number of supersymmetries of the conformal
theory, and N, not italicized, for the dimension of its gauge
group. For a short introduction to string theory and the AdS/CFT
correspondence see, for example, O. Zapata, ``String Theory: A
Theory of Unification,'' [arXiv: hep-th/0612004].}  Even though
the proposal was not well understood by everybody, it was welcomed
and enjoyed rapid acceptance within the community.

\begin{quote}
This subject has developed with breathtaking speed: Maldacena's
paper appeared in November 1997, yet by the Strings 98 conference
seven months later, more than half the invited speakers chose to
speak on this subject.\footnote{ J. H. Schwarz, ``Introduction to
M Theory and AdS/CFT Duality,'' [arXiv: hep-th/9812037].}
\end{quote}

Today, Maldacena's publication is one of the most well-known
papers ever written in theoretical high energy physics.

The first papers submitted to the electronic preprint library
citing Maldacena's conjecture did not delve deeply into the
original proposal; they simply mentioned it in a superficial way.
In these papers we find the following assertions: ``It would be
interesting to understand the relation between our arguments and
those of [reference to Maldacena's paper],'' and ``Maybe an
argument along the lines of [reference to Maldacena's paper] can
be carried out here as well.'' Things changed dramatically when
Edward Witten published a paper formalizing many of the original
ideas put forward by Maldacena.\footnote{ Another important paper
deserving special analysis, what I will do in a next version of
this essay, is S.S. Gubser, I. R. Klebanov and A. M. Polyakov,
``Gauge theory correlators from noncritical string theory,''
[arXiv: hep-th/9802109]. I thank Igor Klebanov for pointing me out
this omission.} He discovered a precise correspondence (here the
term ``correspondence" is used for the first time) between string
states on the ten-dimensional spacetime, dubbed the bulk, and
operators of the particle physics-like model. He also computed
some scattering processes. To this end Witten identified the
boundary of the ten-dimensional bulk with the space where the
non-gravitational particles reside and interact. After this
essential contribution, more and more people started to work on
this correspondence.

Here is how Witten referred to Maldacena's proposal (first lines
of the abstract):

\begin{quote}
Recently, \emph{it has been proposed} by Maldacena that large N
limits of certain conformal field theories in d dimensions can be
described in terms of supergravity (and string theory) on the
product of d + 1-dimensional AdS space with a compact manifold.
Here we elaborate on this idea and propose a precise
correspondence between conformal field theory observables and
those of supergravity.\footnote{ E. Witten, ``Anti de Sitter Space
and Holography,'' [arXiv: hep-th/9802150].} (Italics added.)
\end{quote}

Note Witten's prudence when referring to it: ``It has been
proposed." A year and a half later Maldacena's ``conjecture" was
still a conjecture, that is, nothing exceptional demanded that its
scientific status should be upgraded.  This was the state of
affairs in 1999 when a group of leading string theorists,
including Maldacena himself, published a review article on the
subject. This report comprises more than two hundred and fifty
pages and is still considered one of the most complete accounts on
the subject.

\begin{quote}
So, we conclude that $N$=4 U(N) Yang-Mills theory could be the
same as ten dimensional superstring theory on AdS$_5$ x S$^5$
[reference to Maldacena's paper]. Here we have presented a very
heuristic argument for this equivalence; later we will be more
precise and give \emph{more evidence} for this
correspondence.\footnote{ O . Aharony, S. S. Gubser, J. M.
Maldacena, H. Ooguri and Y. Oz, ``Large N Field Theories, String
Theory and Gravity,'' [arXiv: hep-th/9905111].} (Italics added.)
\end{quote}

In the authors' opinion the correspondence was still in the phase
of gathering evidence. It was not yet established as scientific
fact. At this point it is worth digressing a moment in order to
say a few words about the physics of the correspondence. This will
help us to understand its successive evolution towards a higher
degree of truthfulness.

AdS/CFT in its strongest version states that superstring theory in
the bulk corresponds to a full quantum non-gravitational theory on
the boundary of such volume. But, so far support for it has been
provided only in the supergravity approximation: the point-like
model where the length of the string is equal to zero ($\alpha'
\to 0$). In addition, since these computations are also very
difficult to carry out, the classical limit is necessary. In this
last approximation quantum corrections are discarded (only tree
diagrams are considered). The theory is said to be weakly coupled.

The consistency of the theory relies on every operation being done
on the gravitational theory having a counterpart on the boundary.
(Due to this relationship between what happens in the bulk and on
its boundary, the correspondence has been called holographic.)
Thus, the supergravity and classical limits must have their
corresponding procedure in the boundary theory. Experts have found
that this relationship is of the type weak $\longleftrightarrow$
strong. In short, this means that the easier the computations in
the gravitational theory, as in the limits above, the harder it is
to find the corresponding non-gravitational results on the
conformal field theory side. In turn, when the string calculations
are difficult, the boundary computations are easier to perform.
This explains why the AdS/CFT correspondence is also often called
``AdS/CFT duality.''

After these brief observations, we are now ready to evaluate the
following extract from MAGOO (as the report by Maldacena and
collaborators is sometimes labelled):

\begin{quote}
\emph{One might wonder why the above argument was not a proof
rather than a conjecture.} It is not a proof because we did not
treat the string theory non-perturbatively (not even
non-perturbatively in $\alpha' \to 0$). We could also consider
different forms of the conjecture. In its weakest form the gravity
description would be valid for large $g{_s}$N [supergravity
description with no quantum corrections as explained above], but
the full string theory on AdS might not agree with the field
theory. A not so weak form would say that the conjecture is valid
even for finite $g{_s}$N, but only in the N $\to \infty$ limit (so
that the $\alpha'$ corrections would agree with the field theory,
but the $g_s$ corrections may not). The strong form of the
conjecture, which is the most interesting one and \emph{which we
will assume here}, is that the two theories are exactly the same
for all values of $g_s$ and N.\footnote{ Ibid.} (Italics added.)
\end{quote}

This passage from MAGOO suggests that in those days many string
theorists were not fully convinced of the validity of the
correspondence; although something like 1500 papers had already
been published on the subject (MAGOO includes 757 references in
its bibliography). Despite such wide interest and some important
contributions to theoretical physics, the general opinion was that
the correspondence was in the process of being proved. In their
``Summary and Discussion'' the authors concluded saying:

\begin{quote}
To summarize, the past 18 months have seen much progress in our
understanding of string/M theory compactifications on AdS and
related spaces, and in our understanding of large N field
theories. However, the correspondence is still far from realizing
the hopes that it initially raised, and \emph{much work still
remains to be done}.\footnote{ Ibid.}  (Italics added.)
\end{quote}

So, by May 1999 string theory experts were convinced that the
``simple and powerful observation''\footnote{ I. Klebanov, ``TASI
Lectures: Introduction to the AdS/CFT Correspondence,'' [arXiv:
hep-th/0009139].}  made by Maldacena was in its infancy. At the
same time, an optimistic vision was transmitted to young
researchers by means of courses and written materials. In a widely
used introductory review written by Jens Petersen, which appeared
three months earlier than MAGOO, we read:

\begin{quote}
The Maldacena conjecture [reference to Maldacena's paper] is a
conjecture concerning string theory or M theory on certain
backgrounds of the form AdS$_d$ × M$_{D-d}$. ... The conjecture
asserts that the quantum string- or M-theory on this background is
mathematically equivalent -- or dual as the word goes -- to an
ordinary but conformally invariant quantum field theory in a
space-time of dimension d-1, which in fact has the interpretation
of ``the boundary" of AdS$_d$.\footnote{ J. L. Petersen,
``Introduction to the Maldacena Conjecture on AdS/CFT,'' [arXiv:
hep-th: 9902131].}  (Italics in the original.)
\end{quote}

Similarly, in the written translation of a couple of lectures
delivered during the spring of 1998 at The Abdus Salam
International Centre for Theoretical Physics, two leading
theoretical physicists wrote: ``Assuming this conjecture, one can
derive results for the large 't Hooft coupling limit of gauge
theory, by doing computations in AdS supergravity.''\footnote{ M.
Douglas and S. Randjbar-Daemi, ``Two Lectures on AdS/CFT
Correspondence,'' [arXiv: hep-th/9902022].} And concluded, in the
last lines, by saying: ``Nevertheless, now that we have a precise
and better motivated conjecture for the appropriate string in this
case, we can hope that progress along these lines will be made in
the near future.'' Analysis of written publications at that time
shows that the veracity of the holographic correspondence was
understood by string theoreticians more as a hope rather than as a
completed task.

The lines of development for the following three years, from May
1999 to February 2002, were foreseen, and in a certain sense
dictated, by Maldacena and collaborators: in chapter 5 they
summarized the main results of BTZ black holes and showed how this
was related to the boundary theory; in chapter 6, the final one,
they focused on QCD-like theories.

Thanks to the great amount of available results on the physics of
three-dimensional black holes and a manageable two-dimensional
conformal field theory, the AdS$_3$/CFT$_2$ model was for many
years the favourite setting for analyzing the correspondence
beyond the supergravity limit.

\begin{quote}
In this paper we study the spectrum of critical bosonic string
theory on AdS$_3$ x M with NS-NS backgrounds, where M is a compact
space. Understanding string theory on AdS$_3$ is interesting from
the point of view of the AdS/CFT correspondence since it enables
us to study the correspondence beyond the gravity
approximation.\footnote{ J. Maldacena and H. Ooguri, ``Strings on
AdS$_3$ and the SL(2,R) WZW Model,'' [arXiv: hep-th/0001053]. }
\end{quote}

Juan Maldacena and Hirosi Ooguri continued working on this
framework for some time, arriving at several remarkable results.
Unfortunately, the correspondence was not demonstrated beyond the
supergravity approximation as first expected. The other main line
of research within AdS/CFT was the construction and description of
viable QCD-like theories by means of weak gravitational processes.

\begin{quote}
A fruitful extension of the basic AdS/CFT correspondence
[reference to Maldacena] stems from studying branes at conical
singularities [references]. Consider, for instance, a stack of
D3-branes placed at the apex of a Ricci-flat 6-d cone Y$_6$ whose
base is a 5-d Einstein manifold X$_5$.\footnote{ I. R. Klebanov
and M. J. Strassler, ``Supergravity and a Confining Gauge Theory:
Duality Cascades and $\chi$SB-Resolution of Naked Singularities,''
[arXiv: hep-th/0007191].}
\end{quote}

Even though the two approaches were different, both were trying to
provide evidence for the stronger versions of the correspondence.
The AdS$_3$/CFT$_2$ effort wanted to prove the exact
correspondence in a special case (classical limit with $\alpha'
\neq 0 $), and the AdS/QCD attempt tried to find plausible
phenomenological results. However, by the end of 2001, after four
years of intense work and more than two thousand citations to
Maldacena's original paper, the correspondence was still waiting
for a definitive proof.

The Berenstein-Maldacena-Nastase (BMN) conjecture was proposed in
February 2002 and it rapidly seized the attention of string
theorists working on AdS/CFT. This fervent interest on BMN was
reflected by the large number of publications that followed. In
the month the paper appeared, a fifth of the publications on
AdS/CFT was on BMN or at least mentioned it in the main text. The
following month, articles on BMN grabbed half the attention of the
research on AdS/CFT. A few months later, up four fifths (June and
September 2002) of the citations to Maldacena's 1997 proposal came
from the novel BMN conjecture. This rough count clearly shows that
the new conjecture was an essential breakthrough within the field.
Moreover, as we will see next, it represented the end of a period
and the beginning of a new one. After BMN, the truthfulness of the
AdS/CFT correspondence changed: it was nearing a scientific fact.

In this new conjecture Maldacena and collaborators envisaged an
alternative setting to verify the correctness of the AdS/CFT
correspondence beyond the supergravity limit. The idea was to
concentrate on a very special case of the original formulation and
see how the standard correspondence between string states and
operators matched within this new framework. On the bulk side of
the correspondence the spacetime background was changed to
parallel plane waves. The new condition, pp-waves, was obtained by
taking the Penrose limit of the anti-de Sitter space. In the
conformal field theory this corresponded to a truncation of the
number of operators. It was believed that this new model could
shed light on the full quantum correspondence.

It is interesting to see how Berenstein, Maldacena and Nastase
referred to the AdS/CFT correspondence, the basis of their new
proposal:

\begin{quote}
The \emph{fact} that large N gauge theories \emph{have} a string
theory description was believed for a long time. These strings
live in more than four dimensions. One of the surprising aspects
of AdS/CFT correspondence is the \emph{fact} that for $N$ = 4
super Yang Mills these strings move in ten dimensions and
\emph{are} the usual string of type IIB string theory.\footnote{
D. Berenstein, J. M. Maldacena and H. Nastase, ``Strings in Flat
Space and PP Waves from $N$ = 4 Super Yang Mills,'' [arXiv:
hep-th/0202021].}  (Italics added.)
\end{quote}

These are the first lines of the paper. Such a presentation
suggests that the relationship between gravity and particle
physics is a matter of ``fact.'' They take it for granted.
Obviously, there is something paradoxical in all this: what is
expected to be proven is at the same time considered true
knowledge. But, was this an isolated judgement or rather a belief
shared by other string theorists?

Two months after the BMN proposal, Steven Gubser, Igor Klebanov
and Alexander Polyakov, collaborating then at Princeton, submitted
a paper where it is said:

\begin{quote}
\emph{It was found} in [reference to Maldacena, Witten, and a
previous article by GKP], developing some \emph{earlier findings}
of [reference to Polyakov], that the desired string theory in this
case lives in the space AdS$_5$ x S$^5$ and that \emph{there is a
unique prescription} relating physical quantities in the string
and gauge pictures. Many more complicated examples have been
analyzed since then, \emph{confirming the existence} of a dual
string picture for various gauge theories.\footnote{ S. S. Gubser,
I. R. Klebanov and A. M. Polyakov, ``A Semi-Classical Limit of the
Gauge/String Correspondence,'' [arXiv: hep-th/0204051].}  (Italics
added.)
\end{quote}

Though the authors confess in the next lines that the
correspondence has only been ``tested'' ``mostly in the
supergravity limit,'' as BMN they also presuppose the full
validity of the correspondence. Notice the use of the terms ``it
was found'' and ``confirming the existence.'' The same
predisposition is shown in another important paper written by a
group of researchers from MIT and Harvard:

\begin{quote}
More recently the Maldacena conjecture \emph{has established} a
duality between a conformal gauge theory (with a fixed line of
couplings) and string theories on an AdS background. However these
dualities are well understood only at large values of the gauge
coupling [supergravity limit in the bulk].\footnote{ N. R.
Constable, D. Z. Freedman, M. Headrick, S. Minwalla, L. Motl, A.
Postnikov and W. Skiba, ``PP-Wave String Interactions from
Perturbative Yang-Mills Theory,'' [arXiv: hep-th/0205089].}
(Italics added.)
\end{quote}

A widespread trait among publications following the BMN proposal
is that the few lines making explicit reference to the AdS/CFT
correspondence are often in the abstract or in the first
paragraphs. For instance, the well known article by Joseph Minahan
and Konstantin Zarembo begins with a very short discussion on
AdS/CFT results and limitations. After the six-line review of
AdS/CFT, they move on to the main subject of the paper:
BMN.\footnote{ J. A. Minahan and K. Zarembo, ``The Bethe-Ansatz
for $N$=4 super Yang-Mills,'' [arXiv: hep-th/0212208].} The
function of this brief reference to AdS/CFT in the opening to the
paper is simply to contextualize the article; a context that
everybody must be familiar with, and accept, in order to proceed
further. Post-BMN doctoral dissertations show a similar pattern:
the correspondence is assumed and chapters once intended to
explain it are systematically dropped. A short section or even
several citations now replaced the detailed summary. Another
confirmation that the correspondence was entering a new state
regarding its factuality is that some authors did not even
consider relevant the citation of Maldacena's original paper. From
the eight most important papers\footnote{ I am considering papers
with more than 150 citations on SPIRES (accessed in December
2006).}  on AdS/CFT published after BMN, only four of them cited
it.

What followed in the next years was a confirmation of the previous
analysis. As a sample, let us consider the nine most cited
articles on AdS/CFT during that period.\footnote{ Since it is
impossible for a single person to check the hundreds of papers
written in that period, January 2005 -- December 2006, my
discussion only contemplates the nine most cited papers (each with
more than 50 citations). This segment of the spectrum will show
the main stream of research.}  Three of the papers deal with
phenomenological issues and concentrate on the implications of the
AdS/QCD duality; that is, the possibility of using the holographic
correspondence to obtain precious information on strongly coupled
particle physics processes. As stated in one of these papers:
``Recently, the gravity/gauge, or anti-de Sitter/conformal field
theory (AdS/CFT) correspondence [reference to Maldacena] has
revived the hope that QCD can be reformulated as a solvable string
theory.''\footnote{ J. Erlich, E. Katz, D. T. Son and M. A.
Stephanov, ``QCD and a Holographic Model of Hadrons,'' [arXiv:
hep-ph/0501128].} Another three articles focus on a different
spacetime background for the correspondence, the Lunin-Maldacena
background. This include one written by Oleg Lunin and Juan
Maldacena. The other two are by Sergey Frolov and collaborators:
``A relative simplicity of the Lunin-Maldacena supergravity
background and the $N$=1 superconformal theory makes the
conjectured duality a new promising arena for studying the AdS/CFT
correspondence.''\footnote{ S. Frolov, ``Lax Pair for Strings in
Lunin-Maldacena Background,'' [arXiv: hep-th/0503201].}  In
contrast to the articles on AdS/QCD, these last three are not
phenomenologically motivated; rather they try to prove the
correspondence beyond a constraining condition called the BPS
limit. It is interesting to notice that Lunin and Maldacena called
the new proposal ``conjecture duality,'' while the original
AdS/CFT proposal is simply called ``correspondence.'' This
subtlety differentiation suggests that the latter is in a higher,
better consolidated, factual level. Another of the nine papers on
AdS/CFT concerns integrable models, a subject seeking a solution
to superstring theory on non-trivial backgrounds with RR-fluxes.
There are two more papers. One proposes a sort of AdS/CFT
correspondence for Sasaki-Einstein backgrounds, and the other is
about flux compactifications. Strictly speaking, the last article
is not about the correspondence; it simply acknowledges the
important contribution of the latter to the renewal of the studies
on flux compactifications. And it does it in a single line.

Here concludes our short story of the AdS/CFT correspondence. In
it we saw how string theorists treated the conjecture when it was
proposed for the first time; how they changed their view in the
course of time; and how they communicated it to younger members of
the community. We discovered that the AdS/CFT conjecture became a
fact at the same time as most of the talks and papers changed the
``recently Maldacena conjectured that ...'' to ``as the AdS/CFT
correspondence teaches us ...'' and, finally, to the more
impersonal ``as the AdS/CFT establishes.''  We saw how the
sentence ``Maldacena has recently conjectured that ...''
transformed into a single number that pointed to the original
paper. Nonetheless this was not imposed, as some interpreters
would be incline to declare, by a ``great leader,'' nor by the
``will of power'' of an authoritarian group of researchers, nor by
mere convention. Instead, it is the end result of several years of
long, hard, and exhausting work. I have sustained that the
breaking point was the new ``bold'' conjecture of BMN, a
hypothesis that assumed implicitly the correctness of the old
AdS/CFT correspondence. After years spent accumulating
``evidence," but without a definitive proof in sight, there was
the desire and need within the community to surmount the old
correspondence. The research could safely continue only by
protecting Maldacena's conjecture from profanation, namely,
elevating it to the factual level of the more authentic
mathematical demonstrations. In another context, the historian of
science Steven Shapin wrote: ``It was necessary to speak
confidently of matters of fact because, as the foundations of
proper philosophy, they required protection. And it was proper to
speak confidently of matters of fact, because they were not of
one's own making; they were, in the empiricist model, discovered
rather than invented.''\footnote{ S. Shapin, ``Pump and
Circumstance: Robert Boyle's Literary Technology,'' \emph{Social
Studies of Science} 14, No.4 (Nov., 1984): 481-520.}  To shield
the correspondence from attacks was a necessity for the whole
community of practitioners. Consequently, more and more
discussions on the correspondence were transferred from research
papers and PhD theses to graduate and even undergraduate courses.
This was the final step towards its final entrance into public
lectures and popular science books. Today, the AdS/CFT
correspondence pervades the public debate on superstring theory.

 \end{document}